\documentclass[10pt,twocolumn,floatfix,pra,superscriptaddress]{revtex4-1}
\usepackage{amsmath,amssymb,amsthm,mathrsfs,amsfonts,dsfont,amstext} 
\usepackage{textcomp,pbox}
\usepackage[export]{adjustbox}
\usepackage{bm}
\usepackage{dcolumn,booktabs,url}
\usepackage[scaled]{helvet}
\usepackage{sansmath,gensymb}
\usepackage{tikz,graphicx,transparent,color}
\usepackage{multirow}
\usepackage[separate-uncertainty = true]{siunitx}
\usepackage{comment}
\usepackage{physics}
\usepackage{pdfpages}
\makeatletter
\AtBeginDocument{\let\LS@rot\@undefined}
\makeatother

\usepackage[colorlinks=true]{hyperref}
\usepackage{graphicx}

\hypersetup{
	colorlinks   = true,
	citecolor    = red,
	linkcolor    = blue,
	urlcolor     = red     
}

\graphicspath{{./figs/}}

\newcommand{\yso}{Y$_2$SiO$_5$}
\newcommand{\LN}{LiNbO$_3$}

\newcommand{\ybi}[0]{$^{171}$Yb$^{3+}$}
\newcommand{\ybisotope}[0]{$^{171}$Yb$^{3+}$}
\newcommand{\ybiso}[0]{$^{171}$Yb$^{3+}$:Y$_2$SiO$_5$}

\newcommand{\gstate}{$^2$F$_{7/2}$}
\newcommand{\estate}{$^2$F$_{5/2}$}

\begin{document}

\raggedbottom	%REMOVES UNCESSARY WHITE SPACE BETWEEN PARAGRAPHS

\newcommand{\TitleName}{Remote distribution of non-classical correlations over 1250 modes between a telecom photon and a \ybiso{} crystal}
\title{\TitleName}
	
\newcommand{\AffGeneve}{D\'{e}partment de Physique Appliqu\'{e}e, Universit\'{e} de Gen\`{e}ve, 1205 Gen\`{e}ve, Switzerland}
\newcommand{\AffParis}{Chimie ParisTech, PSL University, CNRS, Institut de Recherche de Chimie Paris, 75005 Paris, France}
\newcommand{\AffPariss}{Facult\'e des Sciences et Ingénierie,  Sorbonne Universit\'e, UFR 933, 75005 Paris, France}

\author{M.~Businger}
\affiliation{\AffGeneve{}}
\author{L.~Nicolas}
\affiliation{\AffGeneve{}}
\author{T.~Sanchez~Mejia}
\affiliation{\AffGeneve{}}
\author{A.~Ferrier}
\affiliation{\AffParis}
\affiliation{\AffPariss}
\author{P.~Goldner}
\affiliation{\AffParis}
\author{Mikael~Afzelius}\email[Email to: ]{mikael.afzelius@unige.ch}
\affiliation{\AffGeneve{}}
	
\date{\today}
	
\begin{abstract}
	Quantum repeaters based on heralded entanglement require quantum nodes that are able to generate multimode quantum correlations between memories and telecommunication photons. The communication rate scales linearly with the number of modes, yet highly multimode quantum storage remains challenging. In this work, we demonstrate an atomic frequency comb quantum memory with a time-domain mode capacity of 1250 modes and a bandwidth of 100~MHz, to our knowledge the largest number of modes stored in the quantum regime. The memory is based on a \yso{} crystal doped with \ybi{} ions, with a memory wavelength of 979~nm. The memory is interfaced with a source of non-degenerate photon pairs at 979 and 1550~nm, bandwidth-matched to the quantum memory. We obtain strong non-classical second-order cross correlations over all modes, for storage times of up to $25~\mu$s. The telecommunication photons propagated through 5 km of fiber before the release of the memory photons, a key capability for quantum repeaters based on heralded entanglement and feed-forward operations. Building on this experiment should allow distribution of entanglement between remote quantum nodes, with enhanced rates owing to the high multimode capacity.	
\end{abstract}

\maketitle 
	
%%%%%%%%%%%%%%%%%%%%%%%%%%%INTRODUCTION%%%%%%%%%%%%%%%%%%%%%%%%%%%%%%%%%%%
%%%%%%%%%%%%%%%%%%%%%%%%%%%%%%%%%%%%%%%%%%%%%%%%%%%%%%%%%%%%%%%%%%%%%%%%%%

\noindent Long-distance networks based on quantum repeaters \cite{Briegel1998} is a key future quantum technology that would allow long-distance quantum key distribution \cite{Duan2001} and other entanglement-based communication tasks between remote quantum systems \cite{Kimble2008,Wehner2018}. Fiber-based quantum repeaters are based on the ability to establish entanglement between stationary nodes (quantum memories) and flying qubits at a telecommunication wavelength (e.g. at 1550 nm) \cite{Simon2007}, which can then be extended over the entire repeater through entanglement swapping \cite{Briegel1998}. Ensemble-based quantum memories are of great interest for quantum repeaters due to their multimode (or multiplexing) capacity, which is crucial for achieving practical rates \cite{Simon2007,Sinclair2014}. The main systems currently under investigation are laser-cooled alkali gases \cite{Pu2017,Parniak2017,Cao2020,Yu2020,Pu2021} and rare-earth ion doped crystals \cite{Askarani2021,Rakonjac2021,lago2021,Liu2021,Ortu2021}.

Solid-state ensemble quantum memories based on rare-earth (RE) ion doped crystals can be multiplexed in time \cite{Simon2007,Afzelius2009a}, frequency \cite{Sinclair2014,Yang2018} and space \cite{Heinze2013,Yang2018,Seri2019}. Time-domain multimode storage is most efficiently achieved with the atomic frequency comb (AFC) quantum memory \cite{Afzelius2009a}, where the mode capacity is proportional to the number of comb lines in the AFC \cite{Afzelius2009a,Ortu2022}. By consequence, a large capacity requires both a large AFC bandwidth and a narrow homogeneous linewidth, i.e. a long optical coherence time, to maximize the number of comb lines. AFC experiments featuring long optical coherence times, in the range of 10 to 100 $\mu$s, have so far only been achieved in nuclear-spin based RE-crystals (non-Kramer ions), namely Eu$^{3+}$:\yso{} \cite{Jobez2016,Ortu2021,Ortu2022}, Pr$^{3+}$:\yso{} \cite{Rakonjac2021,lago2021} and Tm$^{3+}$:Y$_{3}$Ga$_{5}$O$_{12}$ (YGG) \cite{Askarani2021}. However, experiments in Eu$^{3+}$ and Pr$^{3+}$ doped materials have shown AFC bandwidths of around 10 MHz or less, fundamentally limited by the nuclear hyperfine splittings of the same order, which reduces the temporal multimode capacity to the range of 10 to 100 modes \cite{Ortu2022}. Tm$^{3+}$:YGG could potentially store a larger number of modes than Pr$^{3+}$ and Eu$^{3+}$, owing to the enhanced nuclear Zeeman splittings \cite{Thiel2014}. RE ions with electronic spin (Kramer ions), such as Nd$^{3+}$ and Er$^{3+}$, on the other hand, offer higher bandwidths owing to large hyperfine or Zeeman splits, with demonstrated AFC bandwidths ranging from 100 MHz to 6 GHz \cite{Usmani2010,Tiranov2015a,TangZhouWangEtAl2015,Askarani2019,Craiciu2019,Liu2021}. However, the measured AFC coherence times are only in the range of 10 ns to 1 $\mu$s, limited by the increased interaction with the crystal environmenent (e.g. spectral diffusion \cite{Bottger2006a} and superhyperfine coupling \cite{Usmani2010,Craiciu2019}). To maximize the multimode capacity, there is thus a need to find a material where large bandwidth and a long AFC coherence time can be achieved simultaneously.

\begin{figure*}[!t]
\includegraphics[width=\linewidth]{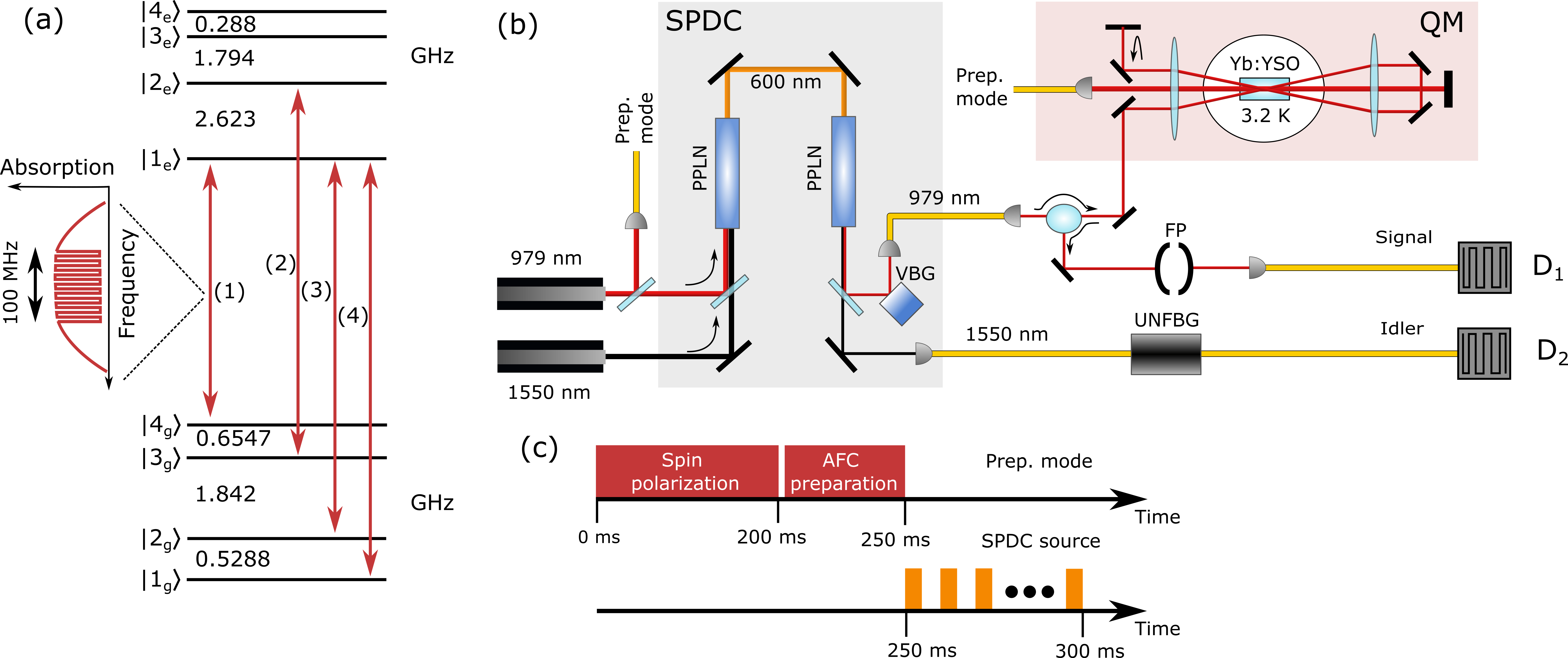}
\caption{\textbf{Experimental setup}. (a) Energy level structure of \ybiso{} at zero magnetic field for site II. The AFC quantum memory is created on transition (1) and transitions (2)-(4) were used for spin polarization into state $\ket{4_g}$. (b) Conceptual experimental setup showing the main components of the SPDC source based on two periodically poled lithium niobate (PPLN) waveguides and the \ybiso{} (Yb:YSO) quantum memory (QM), see main text and Methods for details. To increase absorption, the crystal is mounted in a quadruple-pass configuration, where the backward 979~nm output mode is separated from the input mode via an optical circulator based on a Faraday rotator and a polarization beam splitter. The signal and idler photons were filtered by a volaume Bragg grating (VBG) and Fabry-P\'{e}rot (FP) cavity, and a ultra narrow fiber Bragg grating (UNFBG), respectively, and detected by superconducting nanowire single photon detectors (SNSPDs) D$_1$ and D$_2$. (c) The experimental sequence consists of a spin polarization step (200~ms) and an AFC preparation step (50~ms), as described in the main text. Many storage trials were made during 50 ms, where for each trial the SPDC source was pumped for a duration of $1/\Delta$, followed by detection window of the same duration.}
\label{fig:exp_setup}
\end{figure*}

Here, we report on an AFC quantum memory experiment in \ybiso{} at 979 nm, where 1250 temporal modes are stored owing to its large bandwidth and long optical coherence time. The hyperfine levels in \ybiso{} arise from a highly anisotropic coupling of its electronic spin $S=1/2$  to its nuclear spin $I=1/2$, resulting in four completely hybridized and non-degenerate hyperfine levels in both the excited and ground states \cite{Tiranov2018a}, see Fig.~\ref{fig:exp_setup}(a). At zero magnetic field, these states have zero first-order Zeeman (ZEFOZ) effect, resulting in spin and optical coherence times comparable to those of non-Kramer ions \cite{Ortu2018,Welinski2020}. At the same time, hyperfine splittings range from 288 MHz to 2.623 GHz (for \ybi{} in site II), typical of Kramer ions, and should therefore also allow high-bandwidth quantum memories \cite{businger2020}. In this work, we exploit these unique features to simultaneously demonstrate a large memory bandwidth of 100 MHz and an optical storage time of up to 25 $\mu$s. The memory is interfaced with a photon-pair source based on spontaneous parametric down conversion (SPDC), which produces non-degenerate photons at 979 nm and 1550 nm. Non-classical second-order correlations are obtained for storage times of up to 25 $\mu$s, for which the temporal multimode capacity is shown to be 1250 modes. Furthermore, the telecom photon is propagated through a 5 km fiber spool to demonstrate the ability to establish remote non-classical correlations, a key building block for quantum repeaters. 

\section*{Results}
		
\begin{figure*}[!t]
	\centering
	\includegraphics[width=\linewidth]{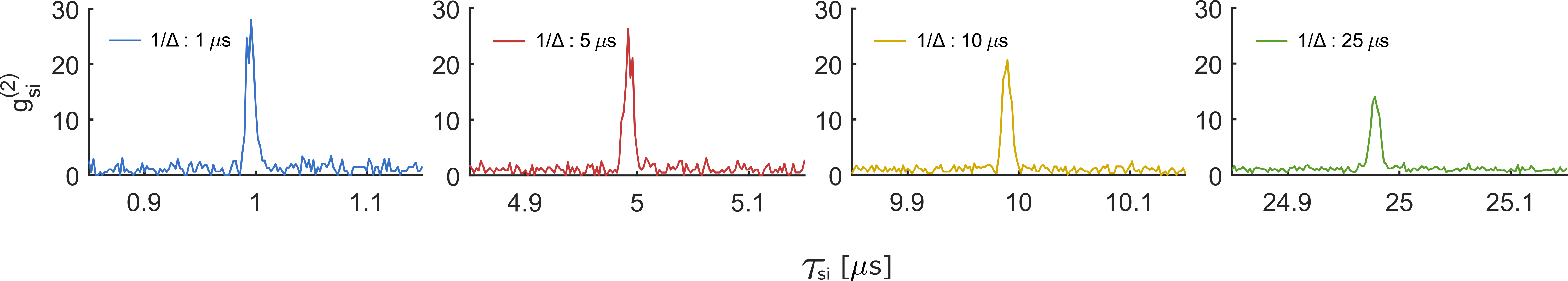}
	\caption{\textbf{Non-classical correlations}. Second-order cross-correlation function $g^{(2)}_{si}(\tau_{si})$ for different AFC storage times 1/$\Delta = $ 1 $\mu$s, 5 $\mu$s, 10 $\mu$s and 25 $\mu$s. The cross-correlation function peaks at the expected emission time of the AFC echo, $\tau_{si} = 1/\Delta $, reaching $g_{si}^{(2)}$($\tau_{si}$)$ = 28(4)$, $26(4)$, $21(2),$ and $ 14(1)$, respectively, well above the classical limit of $2$. All measurements have a binning size of 2 ns. The acquisition times were between 1.15~min for the shortest storage time and up to 8.15~min for the longest storage time.}
	\label{fig:gsidelay}
\end{figure*}

\noindent \textbf{Experimental setup and procedures.} The experimental set-up consists of the \ybiso{} AFC quantum memory and a SPDC source producing correlated photon pairs, see Fig.~\ref{fig:exp_setup}(b). The AFC quantum memory is based on the creation of a comb-like absorption structure in an inhomogeneous absorption profile \cite{Afzelius2009a}, created through spectral hole burning. If the frequency periodicity of the comb is $\Delta$, it leads to a coherent emission of an AFC echo after a pre-determined storage time of $1/\Delta$. In this work, the AFC is created on the optical transition between the lowest crystal-field states of the electronic ground \gstate{} and excited \estate{} states, at a wavelength of 978.854 nm for site II \cite{Welinski2016}, where each electronic state is split into four non-degenerate hyperfine states. The AFC is created on the strong transition connecting $\ket{4_g}$ to $\ket{1_e}$, i.e. transition (1) in Fig.~\ref{fig:exp_setup}(a). To further increase the available optical depth $d$, the experimental sequence starts with a spin polarization step to optically pump ions into the $\ket{4_g}$ state, using transitions (2), (3) and (4), which is then followed by the AFC creation step on transition (1), cf. Fig.~\ref{fig:exp_setup}(c). The \yso{} crystal was doped with 5 ppm of \ybisotope{} \cite{businger2020}, resulting in an inhomogeneous linewidth of 1.3 GHz and an absorption coefficient of about $\alpha = 1.6~\textrm{cm}^{-1}$. The 12~mm long crystal was mounted in a closed-cycle cryostat, cooled to about 3 K, and the 979 nm mode from the SPDC source passed the crystal four times to further increase the optical depth of the AFC (see Fig.\ref{fig:exp_setup}(b).). More spectroscopic properties of this crystal were published in \cite{businger2020}, including population relaxation rates and transition strengths of the optical-hyperfine transitions.

For a finite optical depth $d$ of the AFC transition, the highest echo efficiency is obtained with square-shaped comb teeth, for an optimal comb finesse \cite{Bonarota2010}. Such combs can be efficiently generated by using a complex phase- and amplitude-modulated adiabatic pulse, which simultaneously burns a large number of square-shaped spectral holes at multiple frequencies \cite{Jobez2016}. In Eu$^{3+}$ and Pr$^{3+}$, this preparation method has been successfully implemented to maximize the AFC echo efficiency, while also creating high-resolution AFCs allowing for storage times in the range of 10 to 100 $\mu$s \cite{Jobez2016,Rakonjac2021,lago2021,Ortu2022}. For the low bandwidths in these materials ($<$~10~MHz), this can be achieved using an acousto-optic modulator \cite{Jobez2016}. However, applying the same method for creating large-bandwidth, high-resolution AFCs over 100 MHz, using a high-bandwidth electro-optic modulator (EOM) phase modulator, turned out to be challenging. The periodic, multi-frequency pulse of Ref. \cite{Jobez2016} interferes in the time-domain, resulting in low average power and inefficient optical pumping when the number of frequency bands is increased. For an intensity modulator with fixed maximum modulation amplitude (AOM or EOM), the average power is inversely proportional to the number of comb lines. Here we propose a modified pulse that significantly increases the average pulse energy after the intensity modulator, similarly to the Schroeder method used in telecommunications \cite{Schroeder1970,Oswald2021} (see Supplementary Materials). The pulse allows for efficient creation of a 100 MHz AFC, consisting of up to 2500 comb lines, with a well-defined square shape for each AFC tooth. The same EOM was used to address the (1)-(4) transitions, which spans about 5.6 GHz, by using a combination of arbitrary waveform generators, fixed microwave oscillators and IQ mixers to generate the relevant frequencies, see Methods for details. Note that the spin polarization pulses are also chirped over 100 MHz to achieve a flat initial optical depth before the AFC creation step. 

\begin{figure}[!t]
\includegraphics[width=\linewidth]{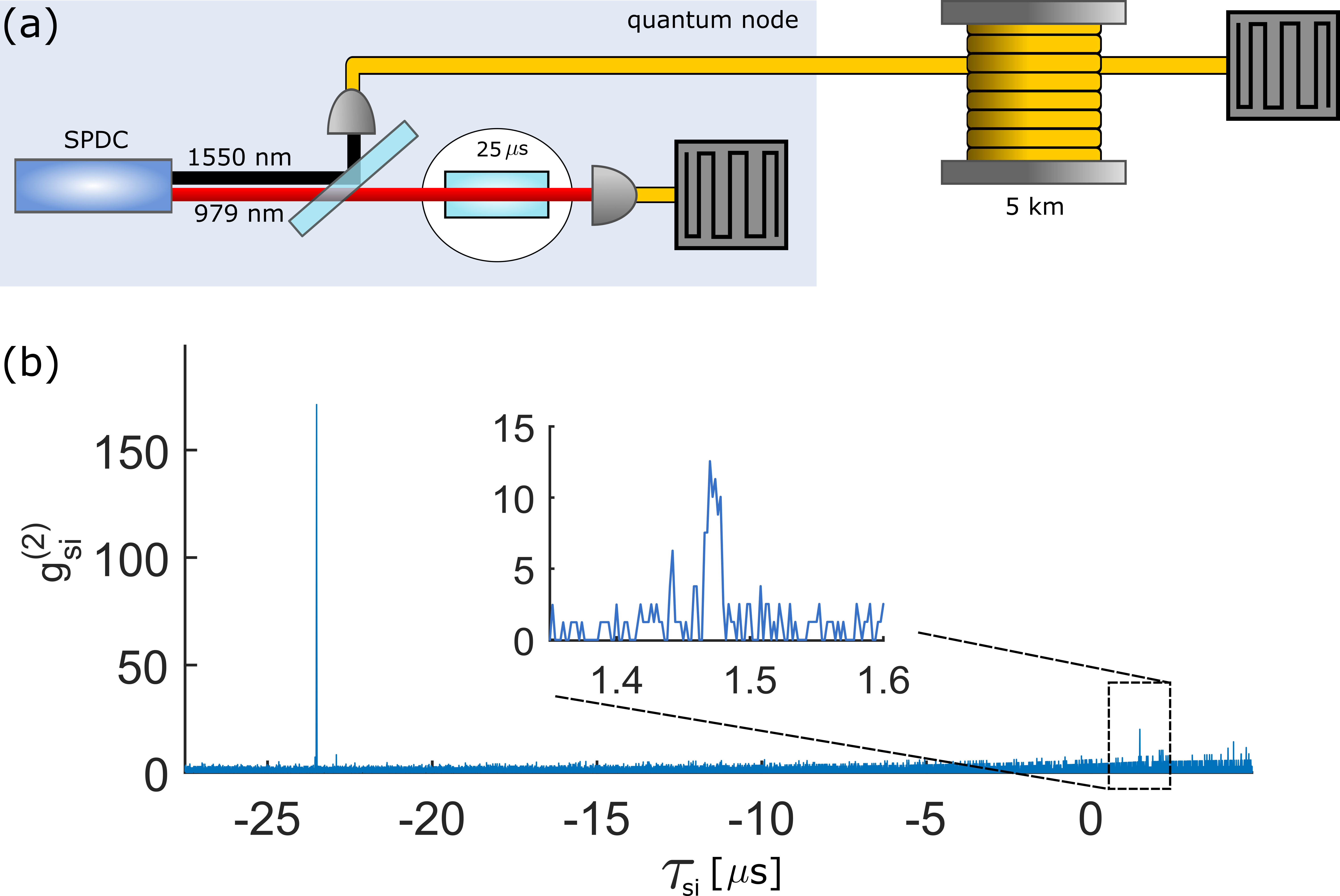}
\caption{\textbf{Remote distribution through a telecom fiber}. (a) A schematic of the storage experiment with a 5-km long fiber spool on the telecom mode, and a memory storage time of 25~$\mu s$. (b) Second-order cross-correlation $g^{(2)}_{si}$ as a function of the relative detection time $\tau_{si}$. The peak at $\tau_{si} = 1.47~\mu$s (see inset) stems from coincidence detections with stored 979~nm photons, whereas the larger peak at $\tau_{si} = -23.5~\mu$s is due to 979~nm photons directly transmitted through the memory. The binning size is 2~ns.}
\label{fig:LongFiber} 
\end{figure}

The SPDC photons were produced by pumping a non-linear waveguide in a \LN{} crystal (PPLN), at a pump wavelength of 600~nm, as shown in Fig.~\ref{fig:exp_setup}(b). The quasi-phase-matching assured that signal-idler photon pairs were produced at 979~nm (signal) and 1550~nm (idler). The pump laser was generated through sum frequency generation (SFG) in an identical PPLN, using the same 979~nm laser which prepares the AFC memory and a telecom 1550~nm laser. To match the bandwidth of the 979~nm photons to the AFC bandwidth, the broad SPDC spectrum (bandwidth of 197~GHz) was filtered down on the signal mode using a Fabry-P\'{e}rot (FP) cavity with a full width at half maximum (FWHM) linewidth of 64~MHz. The idler photons were also filtered down to a FWHM of 500~MHz, to reduce the probability of detecting uncorrelated photons. The filtered 979~nm photons were interfaced with the quantum memory, while the filtered telecom photons were detected directly. Both photons travel through 50 m of fiber before being detected, except in the case of the measurement discussed in relation to Fig.~\ref{fig:LongFiber} where the telecom photon went through a 5-km fiber spool. The timing of the entire experimental sequence is shown in Fig.~\ref{fig:exp_setup}(c). More details on the experimental setup and procedures can be found in the Methods section.

\hfill \break
\noindent \textbf{Non-classical correlations.} From the singles and coincidence detections of the idler and signal photons, one can compute the normalized second-order cross-correlation function $g^{(2)}_{si}(\tau_{si})$ as a function of the relative delay between the detected signal and idler photons $\tau_{si}$. Ideally, the signal and idler modes have thermal statistics individually, in which case the Cauchy-Schwarz inequality implies that $g^{(2)}_{si}(\tau_{si}) > 2$ is a proof of non-classical correlations between the photons \cite{Chou2004}. In an initial characterization, we measured the second-order auto-correlation functions of both modes without the memory in order to reduce losses. These were $g^{(2)}_{ss}(0) = 1.8(1)$ and $g^{(2)}_{ii}(0) = 1.7(2)$, slightly below 2 as expected for thermal states and added detector jitter (see Supplementary Materials). By consequence, $g^{(2)}_{si}(\tau_{si}) = 2$ is a conservative lower bound for non-classical correlations.

In Fig.~\ref{fig:gsidelay}, we show the cross-correlation function $g^{(2)}_{si}(\tau_{si})$ measured for multiple storage times between $1/\Delta =  1~\mu \textrm{s}$ and  $25~\mu \textrm{s}$. The peak cross-correlation value of the emitted AFC echoes after storage are between $g^{(2)}_{si}(\tau_{si}) = 28 $ and 14, demonstrating strong non-classical correlations for all storage times. The decrease in correlation with increasing storage time is mostly due to a drop in memory efficiency and a non-negligible contribution of dark counts on the signal detector D$_1$ \cite{Clausen2014a}.

\begin{figure}[!t]
	\includegraphics[width=1\linewidth]{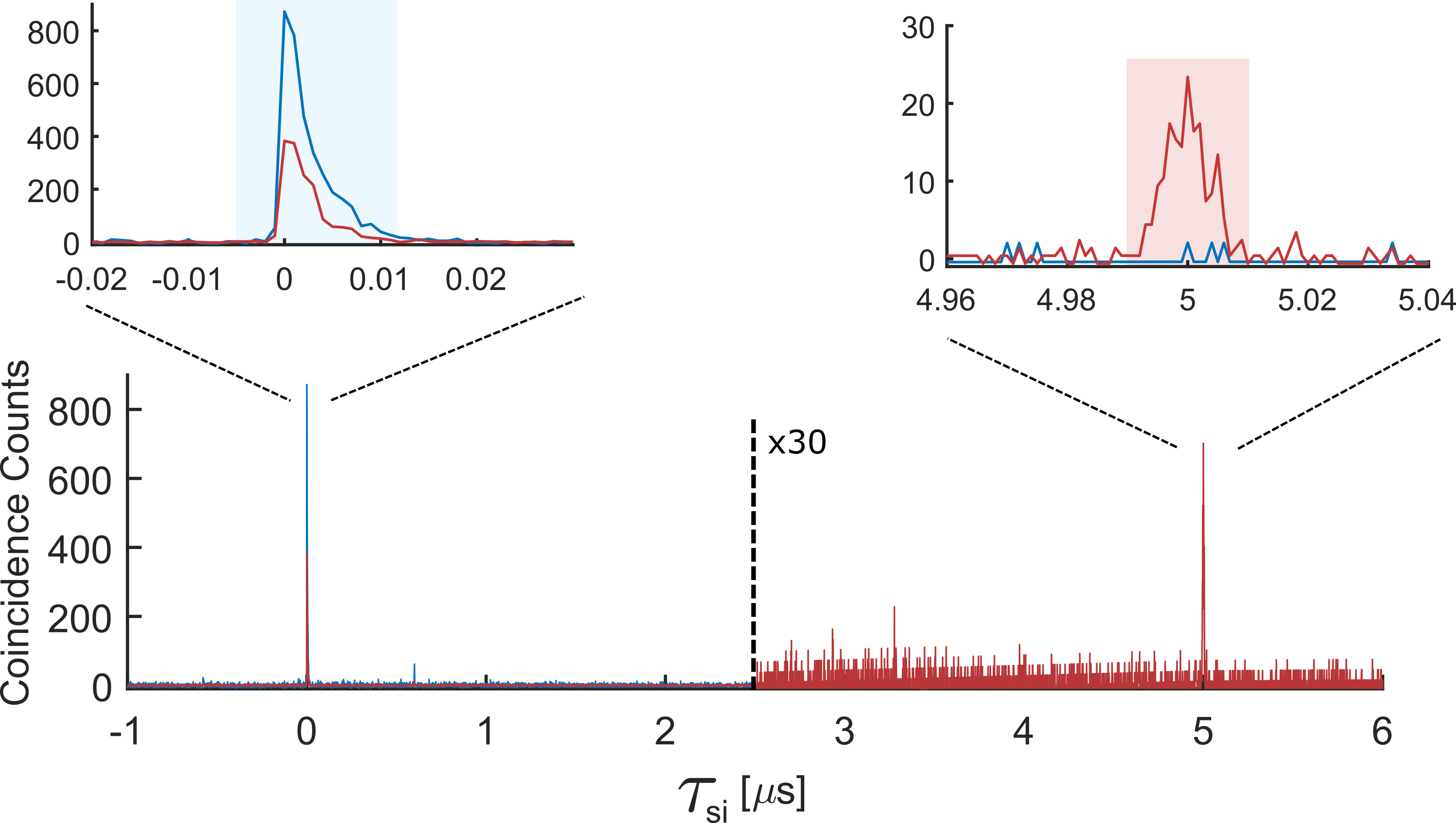}
	\caption{\textbf{Memory efficiency measurement}. Two-photon coincidence detection histograms for an AFC storage experiment with $1/\Delta = 5$ $\mu s$ (red), and for a reference measurement through a transparency window burned into the memory crystal (blue). The shaded areas (see insets) were used to calculate the efficiency, where the echo (reference) integration window was 20 ns (15 ns). Total integration times were 1.15~min (echo) and 0.6~min (reference), respectively. }
	\label{fig:echo_eff}
\end{figure}

\hfill \break		
\noindent \textbf{Memory efficiency and lifetime.} In Fig.~\ref{fig:echo_eff}, we show a coincidence histogram for the storage time of $1/\Delta =  5~\mu \textrm{s}$, together with a reference histogram measured through a transparency window of 200~MHz burned into the inhomogeneous absorption line. The reference histogram allows one to calculate the storage efficiency (see Methods), resulting in an efficiency of $\eta= 4.9 (4)\%$. The efficiency was also characterized as a function of storage time using bright laser pulses (see the Supplementary Materials), which gave $9.8\%$ for $1/\Delta =  5~\mu \textrm{s}$. The difference can be mostly explained by the fact that the 100~MHz wide AFC absorbs at most 60\% of the 64 MHz~wide Lorentzian spectrum of the signal photons. A better bandwidth matching of the memory and signal photons should result in a larger efficiency. We also note that the efficiency is lower than the $15\%$ reached at $1/\Delta =  5~\mu \textrm{s}$ with a 10-MHz wide AFC in the same crystal \cite{businger2020}, which can be mainly attributed to a lower peak optical depth over 100 MHz after the spin polarization step.

The classical efficiency measurement also gave an effective AFC coherence time of $T_2^{\textrm{AFC}} = 69(6)$ $\mu$s. In comparison, up to $T_2^{\textrm{AFC}} = 92(9)$ $\mu$s has been measured in Pr$^{3+}$:\yso{}  \cite{lago2021}, while the longest AFC coherence time of $T_2^{\textrm{AFC}} = 300(30)$ $\mu$s was reached in $^{151}$Eu$^{3+}$:\yso{} \cite{Ortu2021}. Ultimately, the AFC coherence time is limited by the coherence time measured by photon echo, where in \ybiso{} up to  800~$\mu$s was measured \cite{Welinski2020}. In the current experiment, we believe the laser linewidth to be the main limitation of the observed $T_2^{\textrm{AFC}}$ value. 

\begin{figure}[!t]
\includegraphics[width=0.85\linewidth]{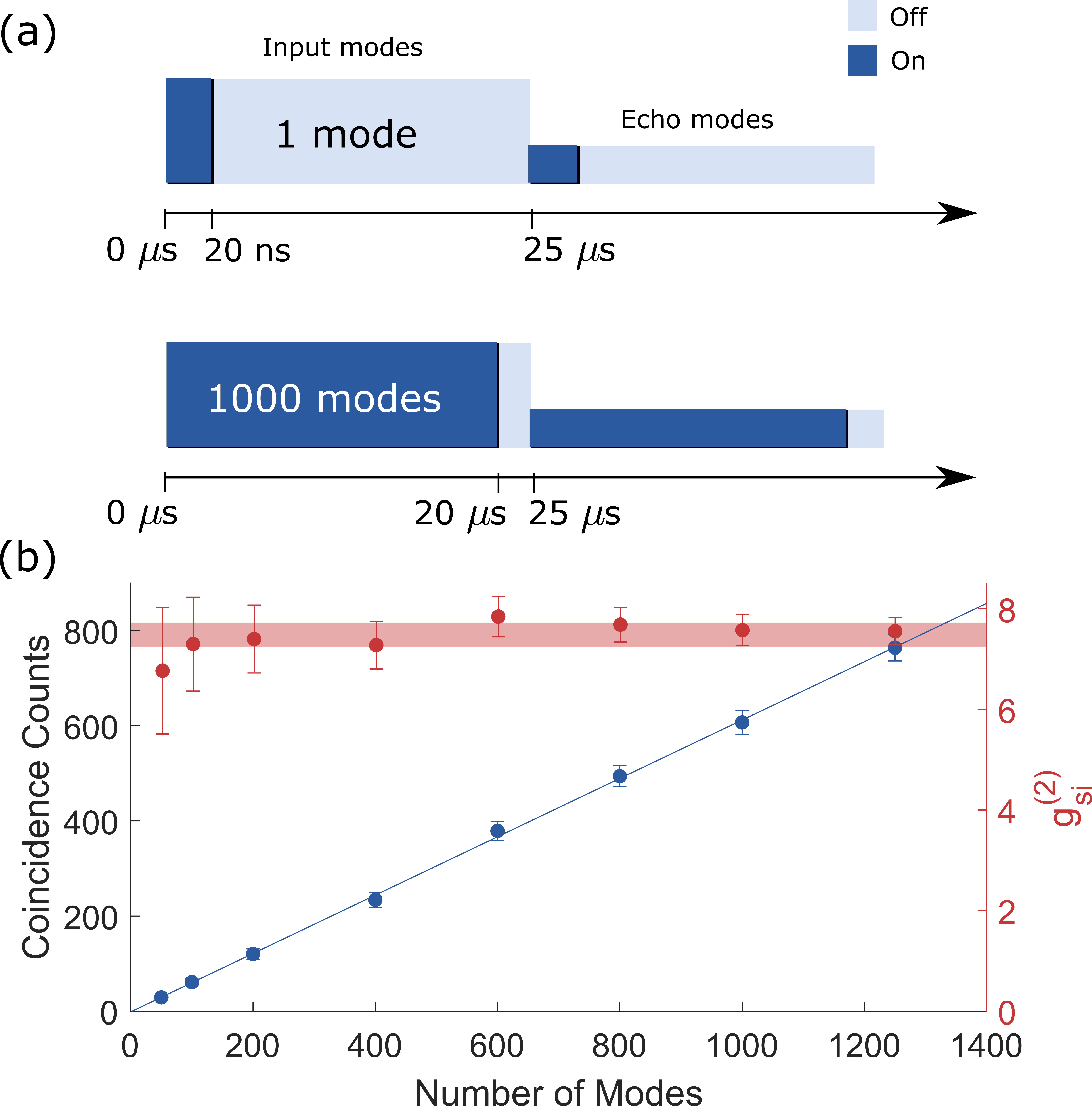}
\caption{\textbf{Temporal multimode capacity}. (a) Concept of the temporal multimode capacity analysis. The storage time is set to $1/\Delta = 25~\mu$s, with each mode having a duration of $T_m = 20$~ns. (b) Total number of coincidence detections and second-order cross-correlation, as a function of the number of analyzed modes. The $g^{(2)}_{si}$ value is calculated by integrating over the entire mode $T_m$, which reduces it to about half of the peak value reported in Fig.~\ref{fig:gsidelay} for 25~$\mu$s.}
\label{fig:Multimode}
\end{figure}

\hfill \break
\noindent \textbf{Quantification of the multimode capacity}. The AFC quantum memory has an inherently large temporal multimode capacity \cite{Afzelius2009a,Nunn2008,Ortu2021} that is key to reaching practical rates in quantum repeaters \cite{Simon2007}. To quantify the multimode capacity, we define a single mode duration of $T_m = 20$~ns that entirely captures the AFC echo histogram peak, cf. Fig.~\ref{fig:echo_eff}.  To show the capability of a multimode memory, with respect to a single mode memory, we analyze the number of coincidence counts and the second-order cross-correlation as a function of the number of temporal modes that are used in post-processing. Similarly to \cite{lago2021}, the time window used for post-processing is progressively increased to change the number of modes used in the analysis. In Fig.~\ref{fig:Multimode}, we show the results of the analysis for the longest storage time of $1/\Delta = 25$ $\mu$s, where we see the coincidence counts increasing linearly with the number of analyzed modes while the cross-correlation stays almost constant at 7.5(2). In total, up to $N = 1/(T_m\Delta) = 1250$ modes were stored while preserving the non-classical correlations, which to our knowledge is the largest number of modes stored in a quantum memory in the non-classical regime.

If we compare to other non-classical storage experiments with at least one telecom photon for remote distribution, Lago-Rivera~\textit{et~al}.~\cite{lago2021} stored 62 temporal modes for $25\mu$s, limited by the much lower AFC bandwidth of about 4~MHz (corresponding to $T_m = 400$~ns) in Pr$^{3+}$:\yso{}. By combining frequency and time modes, Seri~\textit{et~al}.~\cite{Seri2019} stored 135 modes in a laser-written waveguide in Pr$^{3+}$:\yso{}. The development of multimode quantum memories has also progressed quickly in systems based on laser-cooled alkali atoms. Spatial multimode storage using acousto-optic deflectors (AOD) allowed non-classical storage of up to 225 modes in a DLCZ-type quantum memory in $^{87}$Rb, with a memory lifetime of about $28~\mu$s \cite{Pu2017}. Using a single-photon resolving camera Parniak \textit{et al.} \cite{Parniak2017} showed non-classical storage in up to 665 spatial modes for up to $50~\mu$s in $^{87}$Rb, and more recently bipartite entanglement across 500 modes \cite{Lipka2021a}. However, none of these experiments involved a telecom-compatible photon, which would also require interfacing the memory with a quantum frequency converter \cite{Maring_2017a}.

\hfill \break
\textbf{Remote distribution}. With the idler photon in the low-loss telecom window at 1550~nm, our quantum node is ideally suited for remote distribution of non-classical correlations. Given the longest AFC storage time of 25 $\mu s$, and the speed of light in silica fibers, the idler photon can be sent through a fiber spool of 5 km, see Fig.~\ref{fig:LongFiber}(a), and still be detected before the release from the memory of the correlated 979~nm photon. This is a basic requirement for quantum repeaters based on heralded entanglement and feed-forward \cite{Simon2007,Sinclair2014}. In Fig.~\ref{fig:LongFiber}(b), the cross-correlation peak of the emitted AFC echo appears at a relative time delay of $\tau_{si} = 1.47~\mu$s, meaning that the heralding idler detection after the 5 km fiber spool occurs before the 979 nm photon is released from the quantum memory. The $g^{(2)}_{si}(\tau_{si}) = 12(3) $ value is well above the classical limit, only slightly reduced from the short-distance value shown in Fig.~\ref{fig:gsidelay}, showing the capability of the node to distribute non-classical correlations over long distance while locally storing 1250 modes. Building on this experiment should allow distributing entanglement between remote quantum memories as in Ref. \cite{lago2021}, but with a greatly enhanced multimode capacity and correspondingly higher rates.
	
\section*{Discussion}

In this work, we have demonstrated a high multimode capacity, large bandwidth and long optical storage times in \ybiso{}, which are key parameters for ensemble-based quantum repeaters. Yet, these could be improved on in the near future, by exploiting the long intrinsic optical coherence time in \ybiso{} and by extending the bandwidth beyond the current value of 100 MHz. Simulations of the optical pumping process show that at least 250-MHz wide AFCs are possible in principle (for site II), using the same optical transitions as in Ref. \cite{businger2020}. The practical bandwidth is currently limited by the laser power required to efficiently polarize and create the AFC over a large bandwidth, which also limits the memory efficiency. Beyond these technical limitations the memory efficiency can be increased by exploiting cavity enhancement \cite{Moiseev2010a,Afzelius2010a}, which recently was demonstrated with a 500 MHz bandwidth \cite{Davidson2020}.

Other key developments would be to demonstrate on-demand and long-duration storage with the spin-wave AFC technique. In Ref. \cite{businger2020} we showed a 10-MHz bandwidth spin-wave memory with a storage time of 1~ms. A spin-wave bandwidth beyond 100 MHz can realistically be reached, by using a combination of increased laser power in the control fields and exploiting different lambda-systems in the hyperfine manifold with higher control field Rabi frequency (Supplementary Materials).

There is also a highly interesting prospect of interfacing a wide-band \ybiso{} memory with an efficient single photon source based on quantum dots, where recent work has shown that spectrally narrow and indistinguishable single photons can be engineered through cavity-enhanced Raman scattering \cite{Sweeney2014,Beguin2018,BennettAnthony2022}. Such a hybrid solid-state quantum node would pave the way for quantum repeaters with communication rates much higher than possible with current schemes based on probabilistic photon pair sources \cite{Sangouard2007a,Sangouard2011}.

\newpage \clearpage

\section*{Methods}
	
\noindent \textbf{Photon pair source.} The SPDC source consists of two periodically poled lithium niobate (PPLN) waveguides. The first waveguide is used to generate pump photons at 600~nm for the second waveguide that generates correlated photons at 979~nm and 1550~nm. Both waveguides were temperature stabilized using Peltier-elements controlled by a PID. The waveguides have a cross section of 12 $\mu$m $\times$ 10.4 $\mu$m and a length of 3.4 cm. The poling pitch for both crystals is 10.075 $\mu$m. As a pump source for the first waveguide, we use a 1550~nm laser that is amplified to 200 mW and a 979~nm laser at 1 mW that is locked onto the AFC transition. We filter out the pump beam after the first waveguide using two dichroic mirrors and one shortpass filter with a cut off wavelength at 750~nm. After the second PPLN waveguide, a first dichroic mirror picks off the 600~nm light which we use to measure the pump power of the SPDC source. During most measurements, this pump power was kept constant at 270 $\mu$W and only changed for the power study in the Supplementary Material. A second dichroic mirror separates the 979 nm and 1550 nm photons which are both coupled into fibers. The 1550 nm photons are filtered down using an ultra narrow fiber Bragg grating (UNFBG) with a FWHM of 500 MHz which is passively temperature stabilized. We measured the frequency of this filter every 10 min to compensate for any drifts. The 979 nm photons are first filtered by a volume Bragg grating (VBG) with a bandwidth of 20 GHz and, after the memory, by a FP cavity with a FWHM of 64 MHz. The cavity is also temperature stabilized using a Peltier-element controlled by a PID and the central frequency is probed and adjusted every 10 min during the experiments. The cavity has a free spectral range (FSR) of 150 GHz. Additionally, we use a longpass filter at 600 nm before the detector to filter out ambient light from the lab.

\hfill

\noindent \textbf{Experimental sequence.} To prepare the memory, we spin polarize the system into the $\ket{4_g}$ state during 200 ms by driving transitions (2), (3) and (4) as shown in Fig.\ref{fig:exp_setup}(a), where we drive each transition for 1 ms. The AFC is burned on the (1) transition, for 50 ms with 1 ms long complex adiabatic pulses. The pulses are further explained in the following paragraph and in the Supplementary Materials. In addition the preparation mode is gated using an AOM and a fiber based switch which are both synchronized with the 250 ms preparation window, to reduce the burning of the zero order beam and to suppress any leakage of the preparation mode during the detection window. After the spin polarization and AFC preparation steps follows a measurement sequence, where the SPDC source is pumped for a duration of $1/\Delta$ (on period) followed by an off period of the same duration, during which 979 nm photons emitted from the memory can be detected. The total measurement period was 50 ms, where the SPDC source was turned on with a 50\% duty cycle. The SPDC source was gated by a fiber-coupled AOM on the 1550 nm pump light. The entire experimental sequence is being repeated with a repetition rate of 2.3 Hz giving an overall duty cycle of the measurement window of 5.75\%.

\hfill

\noindent \textbf{Memory preparation.} The 979 nm laser used for the preparation mode is frequency stabilized on a home-made, high-finesse cavity and amplified with an optical amplifier. We reach an optical peak power of 40 mW at the memory crystal. The spin polarization pulses are hyperbolic secant (SECH) pulses \cite{Rippe2005} that are programmed with an arbitrary waveform generator and sent into an EOM phase modulator. The pulses have a chirp bandwidth of 100 MHz and a central frequency of 150 MHz. They are up-converted using an IQ mixer with three different local oscillator frequencies, 2.497 GHz, 3.026 GHz and 3.277 GHz, to address transitions (2)-(4) in the preparation sequence. The AFC preparation pulse is also programmed with a bandwidth of 100 MHz and a central frequency of 150 MHz (see Supplementary Information regarding the AFC pulse details). The constant frequency offset of 150 MHz that is applied to all pulses avoids superfluous frequency components in the AFC preparation pulse. The EOM modulation creates high and low frequency sidebands for each modulation frequency, but only the high frequency sidebands are resonant with transitions (1)-(4). The effective peak power of the AFC preparation pulse in the sideband is estimated to be about 10 mW.

\hfill

\noindent \textbf{Efficiency measurement.} To measure the storage efficiency of SPDC photons in our memory, we compare the number of coincidence events that are retrieved after a fixed storage time of 1/$\Delta$ = 5 $\mu$s to the number of coincidence events that would be present without a memory (reference). In the echo window, we measure 173 coincidence events after subtracting accidental coincidences. For the reference measurement we pump a 200 MHz wide transparency window, which had a residual background optical depth of $d_0 = 0.2$. The measured reference coincidences, after subtracting accidental coincidences, was divided by the transmission coefficient $\exp(-d_0)$, yielding 3491 reference coincidences. The estimated efficiency is then $173/3491 \approx 4.9\%$. Fig.~\ref{fig:echo_eff} shows the reference coincidence histogram after subtracting accidentals and compensating for the background optical depth.

\hfill

\noindent \textbf{Data acquisition.} All single photon events shown here were detected with superconducting nanowire single photon detectors (SNSPDs) with efficiencies of 20(5)\% at 979 nm and 80\% at 1550 nm. Background count rates due to dark counts and ambient background light were 30 Hz for the 979 nm detector and 100 Hz for the 1550 detector. All events were recorded using a time to digital converter (TDC) and analyzed in a post processing step. To calculate the coincidences, we used the idler photon as the starting event and the signal photon as the stop event, except for the multimode analysis depicted in Fig.~\ref{fig:Multimode}, where we used the pulse that gates the SPDC source to define the start of the coincidence window such that we were able to distinguish the arrival times of idler photons.

\hfill

\noindent \textbf{Crystal.} The memory is a \ybiso{} crystal with a 5 ppm $^{171}$Yb doping concentration (95\% isotopic purity) that was grown using the Czochralski method. The crystal was cut along orthogonal polarization axes denoted as D$_1$, D$_2$ and b. The dimensions along D$_1$ $\times$ D$_2$ $\times$ b were 3.5 $\times$ 4.5 $\times$ 12 mm, where the light was propagated along the b axis and polarized along D$_2$ to maximize absorption.

\section*{ACKNOWLEDGEMENTS}
We thank S. Welinski and Z. Zhang for interesting discussions and initial work on the crystal growth and characterization, and J. H\"{a}nni for initial work on the photon pair source.
We further acknowledge funding from the Swiss National Science Foundation (SNSF) through project 197168, the European Union Horizon 2020 research and innovation program within the Flagship on Quantum Technologies through GA 820445 (QIA) and GA 820391 (SQUARE), and the French Agence Nationale de la Recherche through project MIRESPIN (Grant No. ANR-19-CE47- 0011).

\bibliography{qmcommon}
	
\end{document}